\documentclass[onecolumn,10pt]{article}

\usepackage{amsmath}
\usepackage{amssymb}
\usepackage{bm}
\usepackage{graphicx}

\begin{document}

\title{Variational approach in dislocation theory}

\author{I.~Groma$^{1}$\footnote{Corresponding author. Email: groma@metal.elte.hu}, G.~Gy\"orgyi$^{2}$\footnote{On leave from: Institute of Theoretical Physics - HAS Research Group, E\"{o}tv\"{o}s University Budapest, Hungary, Email: gyorgyi@glu.elte.hu} and P.D. Isp\'anovity$^{1}$\footnote{Present address: Paul Scherrer Institut, 5232 Villigen PSI, Switzerland. Email: ispanovity@metal.elte.hu}\\
$^1$Department of Materials Physics,
E\"otv\"os University Budapest,\\
P.O. Box 32, 1518 Budapest, Hungary\\
$^2$Department of Theoretical Physics, University of Geneva,\\ 24 quai Ernest Ansermet, 1205 Geneva, Switzerland}
\maketitle

\begin{abstract}
A variational approach is presented to calculate the stress field generated by a system of dislocations. It is shown that in the simplest case, when the material containing the dislocations obeys Hooke's law the variational framework gives the same field equations as Kr\"oner's theory. However, the variational method proposed allows to study many other problems like dislocation core regularisation, role of elastic anharmonicity and dislocation--solute atom interaction. The aim of the paper is to demonstrate that these problems can be handled in a systematic manner.
\end{abstract}

\section{Introduction}
Variational methods are frequently used in many fields of physics. They provide a systematic and compact way to study different problems as well as allow relatively simple ways to incorporate various physical interactions. They make possible to introduce generalised coordinates and variables reflecting the symmetry of the problem. Moreover, they are extremely important in numerical finite element simulations. In dislocation theory, however, variational methods are rarely applied.

One of the few examples is the parametric dislocation dynamics elaborated by Ghoniem \emph{et al.}~\cite{ghoniem}. They have suggested a thermodynamics-based variational method to establish the equations of motion for three-dimensional (3D) interacting dislocation loops. Their approach is appropriate for investigations of plastic deformation at the mesoscopic scale by discrete dislocation dynamics simulations. A fast sum technique for determination of elastic field variables of dislocation ensembles is utilised to calculate forces acting on generalised coordinates of arbitrarily curved loop segments. Each dislocation segment is represented by a parametric space curve of specified shape functions and associated degrees of freedom.

Another example is the Phase Field method proposed by Rodney \emph{et al.}\ to study dislocation dynamics \cite{rodney}. They developed a general formalism for incorporating dislocations in phase field methods based on the elastic equivalence between a dislocation loop and a platelet inclusion of specific stress-free strain. They also discuss how dislocations are elastically and dynamically coupled to any other field such as a concentration field. In their investigations special attention is paid to the treatment of dislocation cores after the discretisation of real and reciprocal space required by the computer implementation of any phase field method. In particular, they propose a method based on two length scales to account for dislocation cores much smaller than the grid spacing.

In this paper we present a variational method that is quite different from the ones mentioned above. It is based on two assumptions:
\begin{itemize}
	\item The total deformation is the sum of the plastic and the elastic deformations, i.e.~we remain in small deformation limit.
	\item The functional form of the Gibbs free energy--stress relation of the material under consideration is assumed to be known.
\end{itemize}
First, the general form of the variational approach is outlined, then we discuss how dislocation core regularisation, elastic anharmonicity and dislocation--solute atom interaction can be treated within the variational framework. 

\section{Derivation of the variational approach}
\label{sec:var}

In order to study the collective properties of dislocations, as a first step, the stress field generated by a discrete dislocation system has to be determined. For a linear elastic medium the problem was formulated in a general way by Kr\"oner \cite{kroner}. In this section we present a variational framework, which is equivalent with Kr\"oner's theory for a medium obeying Hooke's law but it allows to study non-local and non-linear materials too. Elements of this variational approach have been presented in \cite{groma6,groma_anharm}, here we give a detailed derivation and clarify the meaning of the variational functional.

It is commonly assumed that the final state of a body subject to shape changes is reached by two subsequent steps, a plastic and an elastic deformation \cite{Lee}, i.e.
\begin{equation}
\begin{split}
	\partial_j u_i &= \beta_{ij}^\text{t} =(\delta_{ik}+\beta_{ik})(\delta_{kj}+\beta^\text{p}_{kj})-\delta_{ij} \\
	&=\beta_{ij}+\beta^\text{p}_{ij}+\beta_{ik}\beta^\text{p}_{kj},
\end{split}
\label{eq:ep}
\end{equation}
where $u_j$ is the displacement vector, $\beta_{ij}^\text{t}$, $\beta_{ij}$, $\beta^\text{p}_{ij}$ are the total, elastic and plastic distortions, respectively. In our analysis we remain in the small deformation limit, i.e.~we neglect the third term in the right-hand-side of the above equation. Thus the total deformation
\begin{equation}
	\epsilon^\text{t}_{ij}=(\beta^\text{t}_{ij}+\beta^\text{t}_{ji})/2
\end{equation}
is the sum of the elastic $\epsilon_{ij}=(\beta_{ij}+\beta_{ji})/2$ and plastic $\epsilon^\text{p}_{ij}=(\beta^\text{p}_{ij}+\beta^\text{p}_{ji})/2$ deformations:
\begin{equation}
	\epsilon^\text{t}_{ij}=\epsilon_{ij}+\epsilon^\text{p}_{ij}.
\label{epsilon_t}
\end{equation}

The elastic deformation $\epsilon_{ij}$ is generated by the stress field $\sigma_{ij}$. Our aim now is to formulate the condition for the stress in the presence of a given plastic deformation field. For that purpose a convenient way to proceed is to write $\epsilon_{ij}$ as the functional derivative of the Gibbs free energy $G[\sigma]$, functional of the stress tensor components $\sigma_{ij}$, i.e.
\begin{equation}
	\epsilon_{ij}=-\frac{\delta G[\sigma]}{\delta \sigma_{ij}}. \label{H_epsilon}
\end{equation}
With the above relation equation (\ref{epsilon_t}) reads as
\begin{equation}
	\epsilon^\text{t}_{ij}=-\frac{\delta G}{\delta \sigma_{ij}}+\epsilon^\text{p}_{ij}.
\label{epsilon_th}
\end{equation}
We shall use below the incompatibility operation $-e_{ikm}e_{jln} \partial_k\partial_l T_{mn}$ on a tensor $T_{mn}$, where $e_{ikm}$ denotes the antisymmetric unit tensor (sometimes it is defined with opposite sign). Since the incompatibility of $\epsilon^\text{t}_{mn}$ is identically zero, by taking the incompatibility of equation (\ref{epsilon_th}) one obtains
\begin{equation}
	e_{ikm}e_{jln}\partial_k\partial_l\frac{\delta G}{\delta \sigma_{mn}}=\eta_{ij},
\label{H_eta}
\end{equation}
where
\begin{equation}
	\eta_{ij}:=e_{ikm}e_{jln}\partial_k\partial_l\epsilon^\text{p}_{mn}
\end{equation}
is the incompatibility tensor of the plastic deformation. In terms of the dislocation density tensor $\alpha_{ij}$, it has the form \cite{kroner}
\begin{equation}
	\eta_{ij}=\frac{1}{2}\left( e_{jlm}\partial_l\alpha_{im} + e_{ilm}\partial_l\alpha_{jm}\right ).
\label{etaalpha}
\end{equation}
So, we ended up with an equation, (\ref{H_eta}), which relates the stress field to the dislocation distribution and can be used to determine the former if the latter is given.

We should add the condition of equilibrium, $\partial_j \sigma_{ij}=0$,
which is automatically fulfilled by representing the stress tensor as the incompatibility of the stress potential $\chi_{ij}$
\begin{equation}
	-e_{ikm}e_{jln}\partial_k\partial_l\chi_{mn}=\sigma_{ij}.
\label{chi_sigma}
\end{equation}
The functional derivative by the stress potential $\chi$ is easily expressed via partial integration from
\begin{equation}
	\delta G = \int \text{d}V \frac{\delta G}{\delta \sigma_{ij}} \delta \sigma_{ij} =\int \text{d}V \frac{\delta G}{\delta \sigma_{ij}} e_{ikm}e_{jln}\partial_k\partial_l \delta\chi_{mn} =\int \text{d}V \frac{\delta G}{\delta \chi_{ij}} \delta \chi_{ij}
\end{equation}
to yield
\begin{equation}
	e_{ikm}e_{jln}\partial_k\partial_l\frac{\delta G}{\delta \sigma_{mn}}= \frac{\delta G}{\delta \chi_{ij}},
\label{H_chi}
\end{equation}
where $G$ is considered as a functional of the tensor $\chi_{ij}$ only through its dependence on $\sigma_{ij}$ by the definition \eqref{chi_sigma}. Finally, by the use of equation \eqref{H_chi} the equilibrium condition (\ref{H_eta}) becomes
\begin{equation}
	\frac{\delta G}{\delta \chi_{ij}}=-\eta_{ij}.
\label{H_chi_eta}
\end{equation}
By introducing the `plastic free energy' as
\begin{equation}
	P[\chi,\eta]:=G[\chi]+\int \text{d}V \eta_{ij}\chi_{ij},
\label{E}
\end{equation}
equation (\ref{H_chi_eta}) can be reformulated as
\begin{equation}
	\frac{\delta P[\chi,\eta]}{\delta \chi_{ij}}=0.
\label{eq_Emin}
\end{equation}
This way we arrived at a variational principle which yields the equilibrium equations for the stress potential in the presence of dislocations. The functional $P$ has a simple interpretation in that $G[\chi]$ is the elastic Gibbs free energy in the absence of defects, complemented with the term representing the interaction between dislocations and the stress field. Without dislocations, the first term alone is the functional to be extremised, yielding the bulk equilibrium equations. In our formulation the incompatibility tensor assumes the role of the `charge' of dislocations, that interacts with the stress potential linearly, and through the variation by the potential it enters the equilibrium condition.

A further interesting relation can be obtained by taking the plastic free energy (\ref{E}) at the equilibrium stress potential $\chi^\text{eq}$. Substituting equation (\ref{H_eta}) into equation (\ref{E}) and performing partial integrations one concludes that
\begin{equation}
	P[\chi^\text{eq}, \eta]=G[\sigma^\text{eq}]-\int \text{d}V \frac{\delta G[\sigma^\text{eq}]}{\delta \sigma_{ij}^\text{eq}}\sigma_{ij}^\text{eq}, \label{eq_E}
\end{equation}
which is just the Legendre transform of the Gibbs free energy and is thus the elastic free energy of the system as
\begin{equation}
 	F[\eta] = P[\chi^\text{eq}, \eta].
\end{equation}
Note that, before extremisation, $P[\chi,\eta]$ is, in general, not the free energy, but it takes the value of the free energy in elastic equilibrium, when it becomes a functional of the dislocation distribution through the incompatibility tensor $\eta$.

To summarise this section, we can state that we need two constitutive relations to determine the stress field generated by a set of dislocations. The first one is given by equation (\ref{eq:ep}) expressing that the final state of a body subject to plastic deformation is reached by a plastic and a subsequent elastic deformation. In our considerations we neglect the cross term between plastic and elastic deformations, i.e.~we remain in the small deformation limit. The other constitutive relation we need is the Gibbs free energy versus stress functional. With these the stress state generated by a dislocation system can be determined by finding the extremum of the functional given by equation (\ref{E}). It should be mentioned, that the approach explained above provides only the bulk equations. To get a unique solution, one has to set up boundary conditions appropriate for the problem considered which in most cases cannot be naturally obtained from the variational approach proposed. Since in this paper we concentrate only on bulk properties, in the problems explained below we considered either  infinite size systems with vanishing fields at infinity or we used periodic boundary conditions.

Formerly, we have used previous versions of the variational formalism to treat the effect of coarse graining and the correlations it entails, and recognised its use to derive dynamical equations for dislocation densities via the phase field method. Further studies in that direction will be presented in other publications. In the following, we discuss how the extremum principle obtained can be applied in a variety of equilibrium problems, like dislocation core regularisation, studying anharmonic effects and the interaction of dislocations with solute atoms.

\section{Local linear medium}

We first demonstrate the variational principle on the elementary example of a local, linear material obeying Hooke's law. In this case the Gibbs free energy is a quadratic functional of the stress as
\begin{equation}
	G_0[\sigma]:=-\int \frac{1}{2} \sigma_{ij} S_{ijkl} \sigma_{kl} \text{d}V \label{eq:H0},
\end{equation}
where $S_{ijkl}$ is the elastic compliance tensor. Hence, the plastic free energy given by equation (\ref{E}) reads as
\begin{equation}
	P[\chi,\eta]=\int \left[ -\frac{1}{2}e_{iop}e_{jqr}(\partial_o\partial_q\chi_{pr}) S_{ijkl} e_{kst}e_{luz}\partial_s\partial_u\chi_{tz}+\chi_{ij}\eta_{ji} \right]\text{d}V \label{eq_E0}.
\end{equation}
One can find from the equations (\ref{eq_Emin}) and (\ref{eq_E0}) that the stress function fulfils a system of coupled, linear, fourth order partial differential equations
\begin{equation}
	e_{iop}e_{jqr}e_{kst}e_{luz}\partial_s\partial_u\partial_o\partial_q S_{ijkl}\chi_{pr} = \eta_{tz}.
\label{eq_lin}
\end{equation}
Of course, the left-hand-side is just the incompatibility of the elastic deformation tensor. The above equation greatly simplifies for isotropic materials with shear modulus $\mu$ and Poisson's ratio $\nu$, when it becomes (see \cite{kroner})
\begin{equation}
	\triangle ^2 \chi'_{ij}=\eta_{ij},
\end{equation}
where
\begin{equation}
	\chi'_{ij}:=\frac{1}{2\mu}\left(\chi_{ij}-\frac{\nu}{1+2\nu}\chi_{kk}\delta_{ij}\right)
\end{equation}
if $\chi'_{ij}$ satisfies the `gauge condition'
\begin{equation}
	\partial_i\chi'_{ij}=0.
\end{equation}

\subsection{Plane problems}

In the rest of the paper we restrict our analysis to plane (2D) problems, i.e.~we consider only systems of straight dislocations extending parallel to the $z$ direction. After a long but straightforward calculation one can find that for edge dislocations the plastic free energy functional reads as
\begin{equation}
	P[\chi,\alpha]= \int \left[-\frac{1-\nu}{4\mu}(\triangle \chi)^2+\chi(\partial_2 \alpha_{31}-\partial_1 \alpha_{32})\right] \text{d}^2r,
\label{eq_Echi}
\end{equation}
where $\chi:=\chi_{33}$ is now a single component stress function (the other components of $\chi_{ij}$ vanish) and the stress tensor components are
\begin{equation}
	\sigma_{11}=-\partial_2\partial_2 \chi, \ \ \sigma_{22}=-\partial_1\partial_1 \chi, \ \ \sigma_{12}=\partial_1\partial_2 \chi.
\label{eq_stress2}
\end{equation}
In the plastic free energy above the incompatibility tensor $\eta_{ij}$ has been expressed by the dislocation density tensor $\alpha_{ij}$, which now has only two components. The extremum condition $\delta P/\delta \chi=0$ leads to the fourth order partial differential equation as
\begin{equation}
	\frac{1-\nu}{2\mu}\triangle^2\chi=\partial_2 \alpha_{31}-\partial_1 \alpha_{32}.
\label{eq_chi}
\end{equation}

Although later in this paper we do not consider screw dislocations, we summarise their case as well. To our knowledge this was the first application of a variational principle in terms of the stress potential by Berdichevsky \cite{berdichevsky}, which motivated us to look for a variational description of general, including edge, dislocations. The plastic free energy now is
\begin{equation}
	P[\phi,\alpha] =\int \left(-\frac{1}{2\mu}|\nabla \phi|^2+\phi\alpha_{33}\right) \text{d}^2r
\label{eq_screw}
\end{equation}
with
\begin{equation}
	\phi:=-\partial_1\chi_{23}+\partial_2 \chi_{31},
\end{equation}
and the relevant stress components are
\begin{equation}
	\sigma_{23}=-\partial_1 \phi, \ \ \sigma_{13}=\partial_2 \phi.
\end{equation}
The corresponding extremum condition gives the Poisson's equation
\begin{equation}
	\frac{1}{\mu}\triangle \phi=-\alpha_{33}.
\end{equation}

It has to be mentioned that the above equations obtained by the variational approach are certainly equivalent with the ones derived earlier by different methods (see Kr\"oner \cite{kroner}). In this section we just demonstrated how the variational method works for a classical local linear medium.

\section{Dislocation core regularisation}

The significance of dislocation core regularisation is widely known. It is not only necessary for the explanation of physical core effects, but also to eliminate singularities in a physically well founded manner in numerical simulations. There are many different propositions for dislocation core regularisation (for a review see \cite{aifantis}) but, as it is explained below, the variational approach offers a natural way to regularise the singular stress at the dislocation line.

It is common in phase field theories that surface or size effects are captured by introducing appropriate `gradient terms' in the energy functional. The concept can be applied in dislocation theory too, but as we have recognised above, the physical properties of a material are determined by the functional form of the Gibbs free energy--stress relation. So, the gradient terms should be introduced in the Gibbs free energy and we have to add terms that depend on the gradient of the stress. In first order linear approximation one can consider the `non-local' Gibbs free energy
\begin{equation}
	G_\text{non-local}[\sigma] := G_0[\sigma] - b^2\int N_{ijklmn}(\partial_i \sigma_{jk})\partial_l \sigma_{mn} \text{d}V,
\label{eq_nonlocal}
\end{equation}
where $N_{ijklmn}$ is a constant tensor with inverse stress dimension and $b$ is the Burgers vector. ($b^2$ is separated from $N_{ijklmn}$ to indicate the relative order between $G_0$ and the gradient dependent term.) From $G_\text{non-local}[\sigma]$ the corresponding $P[\chi,\eta]$ has to be constructed as it is explained in section \ref{sec:var}.

It should be mentioned that non-locality could be introduced on a much more general way by taking the Gibbs free energy in the form
\begin{equation}
	G_\text{non-local}[\sigma] := -\int \frac{1}{2} \sigma_{ij}({\bm r}) S_{ijkl}({\bm r}-{\bm r}') \sigma_{kl}({\bm r}') \text{d}V\text{d}V'
\end{equation}
where $S_{ijkl}({\bm r})$ is a function which goes to zero fast enough if $|{\bm r}|\rightarrow \infty$, but in a first order approximation, if its range is of the order of the lattice constant, it obviously gives the same as (\ref{eq_nonlocal}).

\begin{figure}
\begin{center}
\includegraphics[width=7cm, angle=-90]{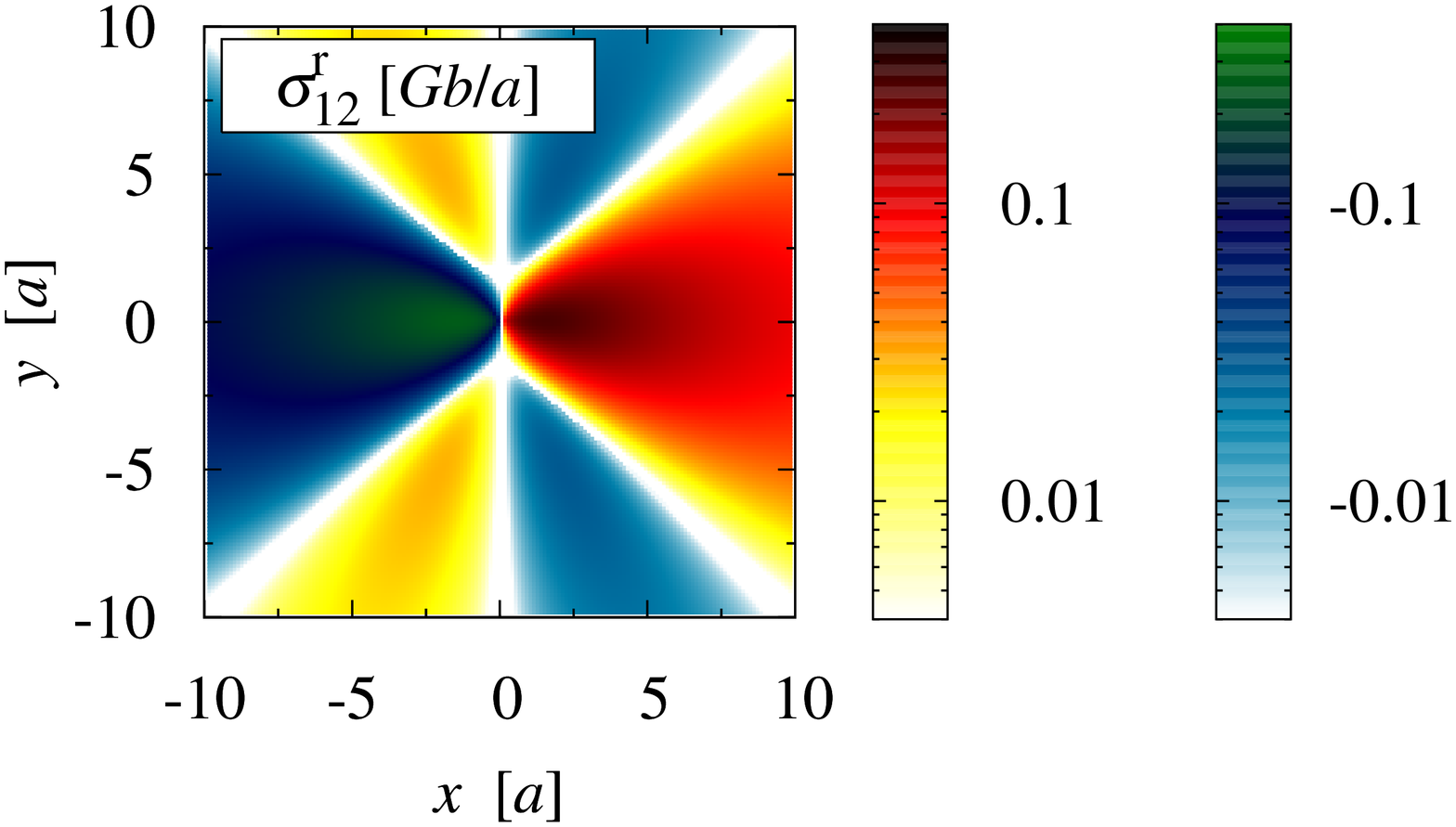} \\
\includegraphics[width=7cm, angle=-90]{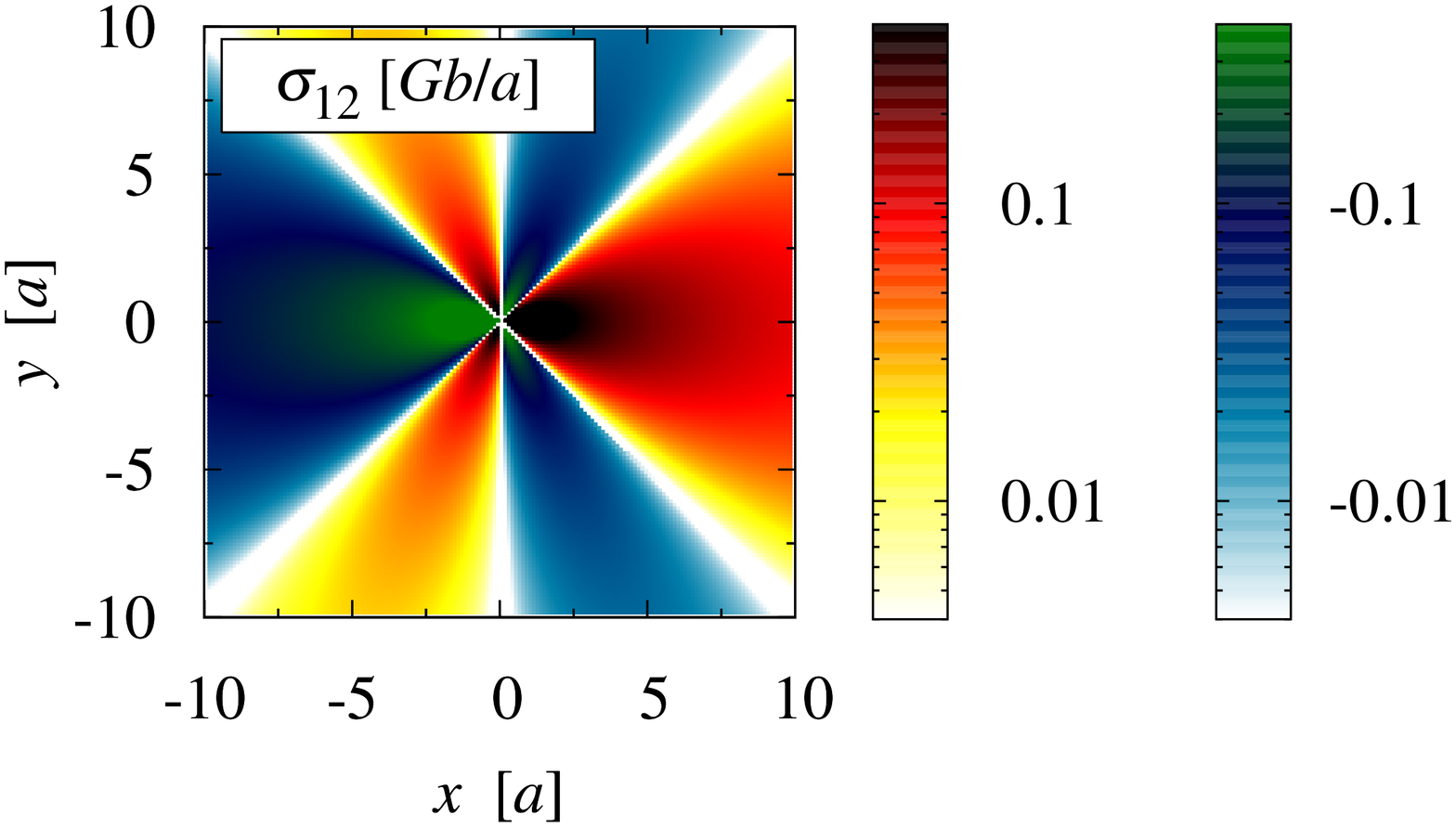}
\end{center}
\caption{Shear stress obtained with (upper box) and without (bottom box) core regularisation. Here the $G:= \mu / [2 \pi (1-\nu)]$ elastic constant was introduced. \label{fig:tau_eff}}
\end{figure}

\begin{figure}
\begin{center}
\includegraphics[width=8cm, angle=-90]{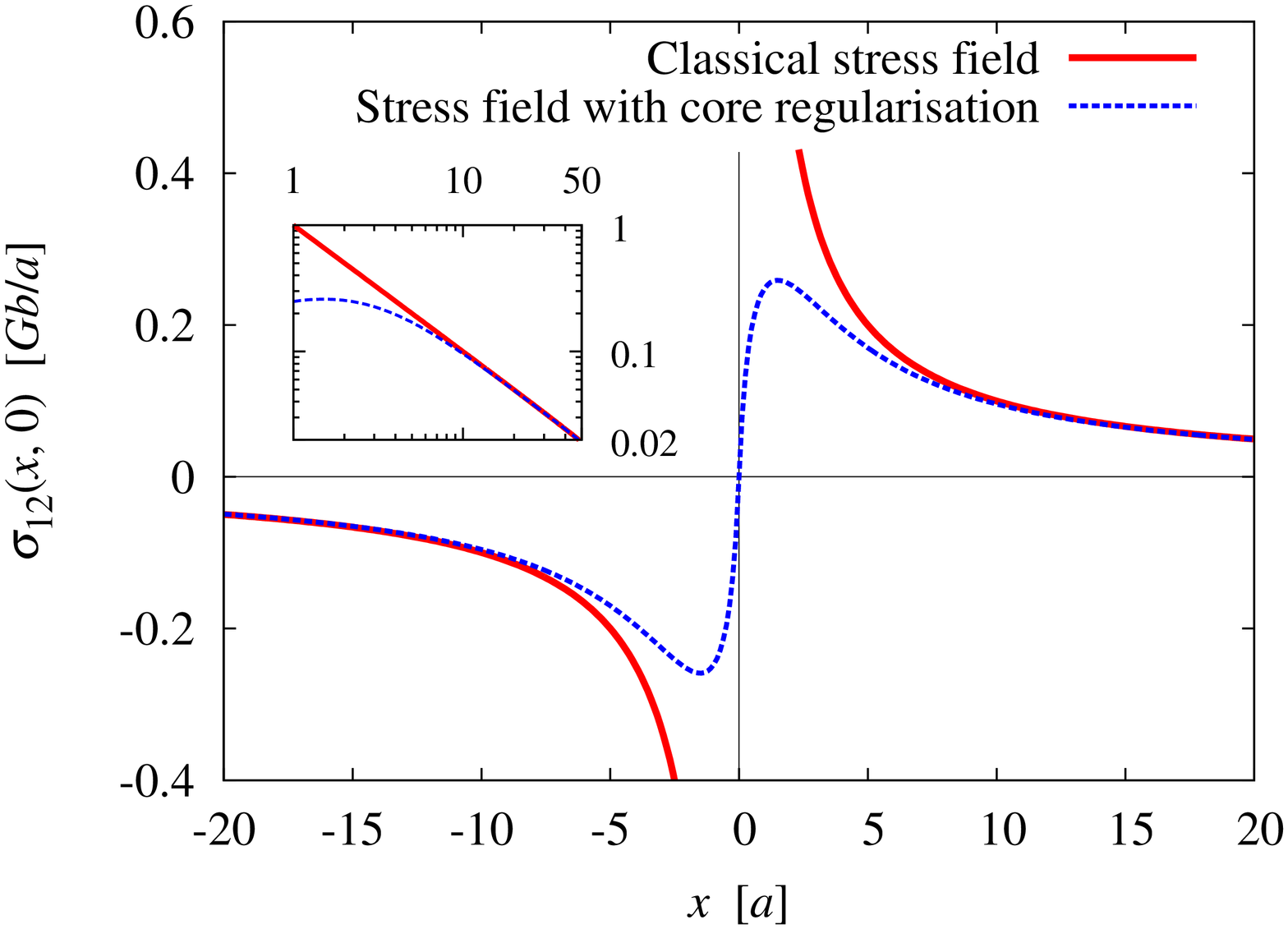}
\end{center}
\caption{Shear stress along the $x$ axis obtained with and without core regularisation. ($G:= \mu / [2 \pi (1-\nu)]$.) \label{fig:tau_x}}
\end{figure}

To demonstrate how the non-local term introduced above results in dislocation core regularisation let us consider a single straight dislocation embedded in an isotropic medium. From equations (\ref{eq_Echi}) and (\ref{eq_nonlocal}) for a single edge dislocation at the origin
\begin{equation}
	P[\chi]= \int \left\{-\frac{1-\nu}{4\mu}\left[(\triangle \chi)^2+a^2|\nabla\triangle \chi|^2\right]+\chi\partial_2 \delta({\bm r})\right\} \text{d}^2r, \label{eq_chi_non}
\end{equation}
where $a$ is a parameter with length dimension that is in the order of the lattice constant. Here, for the sake of simplicity, we considered only the simplest possible isotropic gradient term from \eqref{eq_nonlocal} but the general case can be treated in a similar way. The corresponding equilibrium equation has the form
\begin{equation}
	\triangle^2\chi-a^2\triangle^3\chi=\frac{2b\mu}{1-\nu}\partial_2 \delta({\bm r}).
\end{equation}
The above equation can be solved numerically quite easily. By taking its Fourier transform one can find that
\begin{equation}
	\chi^F(q_1,q_2)=\frac{2b\mu}{1-\nu}\frac{iq_y}{(q_x^2+q_y^2)^4+a^2(q_x^2+q_y^2)^6}
\end{equation}
from which, according to equation (\ref{eq_stress2}), the Fourier transform of the resolved shear stress reads as
\begin{equation}
	\sigma^{\text{r},F}_{12}(q_1,q_2)=-\frac{2b\mu}{1-\nu}\frac{iq_xq_y^2}{(q_x^2+q_y^2)^4+a^2(q_x^2+q_y^2)^6}.
\end{equation}
By numerical inverse Fourier transformation one obtains the shear stress plotted in figure \ref{fig:tau_eff}.
To demonstrate the difference more explicitly the shear stress along the $x$ axis is plotted in figure \ref{fig:tau_x}.
As it can be seen, for $x$ larger than about $10\,a$, $\sigma^\text{r}_{12}$ is close to the classical stress field $\sigma_{12}\propto 1/r$. So, as it is expected, the `gradient term' introduced above influences only the central core region.

In a similar way, for screw dislocations from equations (\ref{eq_screw}) and (\ref{eq_nonlocal})
\begin{equation}
	P =\int \left\{-\frac{1}{2\mu}\left[|\nabla\phi|^2+c^2(\triangle\phi)^2 \right] +b\phi\delta({\bm r})\right\} \text{d}^2r, \label{eq_screw_non}
\end{equation}
where $c$ is a constant. Such approach was also made in \cite{berdichevsky}. One can obtain that the extremum condition $\delta P/\delta \phi=0$ is fulfilled if $\phi$ satisfies the
equation
\begin{equation}
	\triangle \phi-c^2\triangle ^2\phi=-b\mu \delta({\bm r}).
\end{equation}
Like for edge dislocations, numerical solution of the above equation shows that the second term of the left hand side of the equation results in stress regularisation in the vicinity of the dislocation line.

To summarise this section, we have shown that by adding appropriate terms, depending on the gradient of the stress tensor, to the Gibbs free energy functional one can regularise the dislocation core. We have discussed in detail only the simplest possible `isotropic' core regularisation, but one can study more general cases too. Comparing the present approach with other methods suggested by Aifantis \emph{et al.}~\cite{aifantis,gutkin}, and Lazar \cite{lazar}, although one can find some similarities, but the major difference is that in these works the core region is regularised by spreading out the dislocation density tensor $\alpha_{ij}$, while in our analysis the dislocation density tensor remains proportional to a Dirac delta, like in the classical non-regularised case.

Finally, it should be stressed, that the physical justification of the results obtained above requires further detailed investigations. This is beyond the scope of this paper, our aim was only to demonstrate that the problem can be treated within the variational approach in a straightforward manner.

\section{Role of elastic anharmonicity}

There is quite an extended literature  on how to deal with dislocations in the geometrically and materially non-linear setting \cite{kroner_seger,gariola,malyshev}. Among other things the stress field of  single straight edge and screw dislocations are analytically determined. So far, however, the interaction energy of dislocations is not calculated for the non-linear case. Recently, Nabarro and Brown have suggested that the energy difference between dipoles with interstitial and vacancy types can play an important role in the physical properties of PSBs \cite{brown1,brown2}. As it was realised by them, in order to account for this energy difference one should go beyond linear elasticity. Allowing a certain quadratic term in the stress--strain relation they were numerically able to estimate the energy difference between the two types of dipoles\cite{brown2}. In this section we show that the problem can be treated in a systematic manner within the variational approach explained in section 2.

Before we start our analysis, we have to compare the different possible sources of non-linearities in dislocation theory. As it is discussed at the beginning of section 2, according to equation (\ref{eq:ep}), the general form of the total distortion contains the product of elastic and plastic distortions. This term was neglected in the subsequent considerations. However, as it is explained below, elastic anharmonicity plays a much more important role. Up to quadratic terms anharmonicity can be accounted by introducing a quadratic term in the stress--strain relation:
\begin{equation}
	\sigma_{ij}=L_{ijkl} \epsilon_{kl}+K_{ijklmn} \epsilon_{kl}\epsilon_{mn},
\end{equation}
where $L_{ijkl}$ and $K_{ijklmn}$ are the first and second order elastic moduli, respectively.
According to the MD simulations of Chantasiriwan and Milstein \cite{chanta}, for cubic metals the ratio of the first and second order moduli is $\delta=K_{1111}/L_{11}\approx -10$. ($\delta= -9.8$ was reported experimentally on steel by Sommer \emph{et al}.~\cite{sommer}.) In contrast with this, in equation (\ref{eq:ep}) the `coupling constant' between the linear and the quadratic terms is $0.5$. Due to this, the non-linear effects related to elastic anharmonicity are an order of magnitude higher than the ones connected to the other non-linearity mentioned above. So, the assumptions needed to derive the variational framework are justified.

Since in the variational method proposed the Gibbs free energy--stress relation represents the physical properties of the material under consideration, anharmonicity has to be taken into account by adding cubic and/or higher order terms to the quadratic $G_0[\sigma]$ relation given by equation (\ref{eq:H0}):
\begin{equation}
	G_{\rm anharm}[\sigma] := G_0[\sigma]+\int \left[C_{ijklmn}\sigma_{ij}\sigma_{kl}\sigma_{mn}+ {\cal O}(\sigma_{ij}^4) \right] \text{d}V, \label{anharm}
\end{equation}
where $C_{ijklmn}$ is a constant tensor of rank six. In the following, we briefly show that the anharmonicity leads to an extra term in the interaction energy of two parallel edge dislocations. The reader can find the detailed derivation in \cite{groma_anharm}. For simplicity we consider isotropic medium and take the following Gibbs free energy
\begin{equation}
	G[\chi]:=\int \left[-\frac{D}{2}(\triangle \chi)^2 +\epsilon_\text{a}\frac{D^2}{3}(\triangle \chi)^3 \right ]\text{d}^2r, \label{2d_H1}
\end{equation}
where $D:=(1-\nu)/2\mu$ and $\epsilon_\text{a}>0$ is a dimensionless coupling parameter. Since, according to equation (\ref{eq_stress2}), $\triangle \chi=-(\sigma_{11}+\sigma_{22})=2p$, the anharmonic term given above is proportional to the third power of the pressure $p$. So, it represents `inclusion--exclusion' asymmetry. One can find that the other possible cubic terms in (\ref{anharm}) result in quite similar dislocation--dislocation interactions than the one considered here.

The corresponding plastic free energy defined by equation (\ref{E}) is
\begin{equation}
	P[\chi]=\int \left\{-\frac{D}{2}(\triangle \chi)^2 +\epsilon_\text{a}\frac{D^2}{3}(\triangle \chi)^3 + \chi[b_1\partial_y \delta({\bm r} -{\bm r}_1)+b_2\partial_y \delta({\bm r} -{\bm r}_2)] \right \}\text{d}^2r, \label{2d_Ea}
\end{equation}
where ${\bm b}_1=(b_1,0,0)$ and ${\bm b}_2=(b_2,0,0)$ are the Burgers vectors ($|b_1| = |b_2| =: b$), and ${\bm r}_1$ and ${\bm r}_2$ are the positions of the two dislocations, considered as fixed for now.

In order to calculate the interaction energy between two dislocations, we make use of the property that the plastic free energy extremised by the stress function will assume the value of the free energy. So, if we start out from the plastic free energy with two dislocations in fixed positions, extremise, then substitute the equilibrium stress function, the term that depends on both positions will be just the sought interaction energy. While the leading pair energy from linear elasticity is long known, the anharmonic correction will be determined as follows.

First, we have to calculate the stress function $\chi({\bm r},{\bm r}_1,{\bm r}_2)$ generated by two dislocations. One obtains from equations (\ref{eq_Emin}) and (\ref{2d_Ea}) that
\begin{equation}
	D\triangle^2 \chi-\epsilon_\text{a}D^2\triangle \left[(\triangle \chi)^2 \right] = b_1\partial_y \delta({\bm r} -{\bm r}_1)+b_2\partial_y \delta({\bm r} -{\bm r}_2).
\label{chi_2}
\end{equation}
Since $\epsilon_\text{a}$ is a small parameter, up to first order in $\epsilon_\text{a}$, the solution of equation (\ref{chi_2}) can be found as
\begin{equation}
	\chi({\bm r},{\bm r}_1,{\bm r}_2)= \, \chi_1({\bm r} -{\bm r}_1)+\chi_2({\bm r} -{\bm r}_2) +\epsilon_\text{a}\xi({\bm r},{\bm r}_1,{\bm r}_2),
\label{chi_1}
\end{equation}
where
\begin{equation}
	\chi_i({\bm r}):=\frac{b_i}{b}\chi_0({\bm r}), \hspace{15pt} i=1,2 \label{chi12}
\end{equation}
is the solution of the harmonic single dislocation problem
\begin{equation}
	D\triangle^2 \chi_0= b\partial_y \delta({\bm r}),
\end{equation}
\begin{equation}
	\chi_0(\bm r) = \frac{b}{8\pi D} \partial_y r^2 \ln(r). \label{chi_0e}
\end{equation}
After substituting equations (\ref{chi_1}) and (\ref{chi_0e}) into equation (\ref{chi_2})
one can find that if $\epsilon_\text{a}$ is small than $\xi$ satisfies the equation
\begin{equation}
	\triangle\xi=D(\triangle \chi_1+\triangle\chi_2)^2. \label{laplace_xi}
\end{equation}

According to the general expression (\ref{eq_E}), the plastic free energy (\ref{2d_Ea}) at the extremising stress function gives the free energy. Substituting equation (\ref{chi_2}) into (\ref{2d_Ea}) gives
\begin{equation}
	F = P[\chi]=\int \left[\frac{D}{2}(\triangle \chi)^2 -\epsilon_\text{a}\frac{2D^2}{3}(\triangle \chi)^3 \right ]\text{d}^2r, \label{2d_E1}
\end{equation}
wherein we now understand the equilibrium $\chi$ field. Up to terms linear in the coupling parameter $\epsilon_\text{a}$, the total free energy is
\begin{equation}
	F=F_0+\epsilon_\text{a}F_1,
\end{equation}
where $F_0$ is the well-known energy in harmonic approximation and
\begin{equation}
	F_1:=\int \left\{ D \triangle (\chi_1+\chi_2) \triangle \xi-\frac{2D^2}{3}[ \triangle (\chi_1+ \chi_2)]^3 \right \} \text{d}^2r. \label{E_1}
\end{equation}
From equations (\ref{laplace_xi}) and (\ref{E_1})
\begin{equation}
	F_1=\int \frac{D^2}{3}[ \triangle (\chi_1+ \chi_2)]^3 \text{d}^2r. \label{E_1_2}
\end{equation}
This can be rewritten as
\begin{equation}
	F_1=D^2\int [ (\triangle \chi_1) (\triangle \chi_2)^2+(\triangle \chi_2) (\triangle \chi_1)^2] \text{d}^2r +C, \label{E_1_3}
\end{equation}
where $C$ is a constant representing the correction to the single dislocation energy. While this is singular in the present approximation, it does not contribute to the dislocation--dislocation interaction force. Thus, the $V_1({\bm r}):=F_1-C$ interaction energy correction has the form
\begin{equation}
	V_1({\bm r})= \frac{(b_1-b_2)b^2}{(2\pi)^3D}\int \frac{y'-y}{|{\bm r}'-{\bm r}|^2}\left(\frac{y'}{|{\bm r}'|^2}\right)^2 \text{d}^2 r', \label{V1}
\end{equation}
where ${\bm r}:={\bm r}_1-{\bm r}_2$ is the relative coordinate of the two dislocations.

The integral in the above energy correction can be calculated analytically. Since it is not straightforward, we give the detailed calculation in the Appendix. One finds that
\begin{equation}
	V_1(\bm r) = \frac{(b_1-b_2)b^2}{16\pi^2 D} \left[ \frac{y(3x^2+y^2)}{r^4} + 4 \frac{y}{r^2} \ln\left( \frac{r}{r_0} \right) \right].
\label{V1_f}
\end{equation}
where $r_0$ is the inner cut-off radius. Please notice that a term is missing from the energy correction form (and so the extra interaction force) given in \cite{groma_anharm}.

It is useful to summarise some of the properties of $V_1$:
\begin{itemize}
\item If the Burgers vectors of the two dislocations are the same, $V_1$ vanishes, so the non-linearity considered generates extra interaction between dislocations of parallel Burgers vectors only if their signs are opposite.
\item $V_1$ decays as $\ln(r)/r$ with a complicated angular dependence.
\item $V_1$ is antisymmetric in $y$.
\item $V_1$ is symmetric in $x$.
\item If we take a dislocation and then we put another one with opposite sign above or below, the interaction energy of the two configurations are not the same due to the extra interaction term $V_1$. So, as it is expected, the anharmonicity introduced recovers the well know fact that the interaction energy of the interstitial and vacancy type dislocation dipoles are not the same.
\end{itemize}

A remarkable consequence of the extra term in the dislocation--dislocation interaction energy is that it strongly modifies the relaxed dislocation morphology developing from an initially random dislocation configuration.
\begin{figure}
\begin{center}
\includegraphics[width=5.5cm,angle=-90]{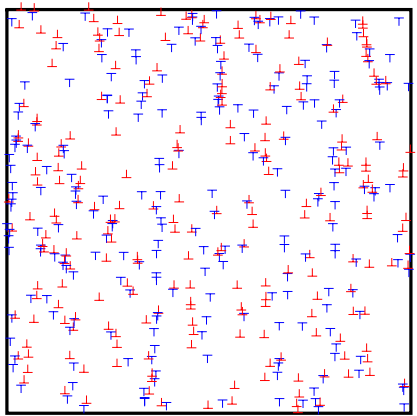}
\end{center}
\caption{A relaxed dislocation configuration with $\epsilon_\text{a}=0$ (harmonic case). \label{fig:harm}}
\end{figure}

\begin{figure}
\begin{center}
\includegraphics[width=5.5cm,angle=-90]{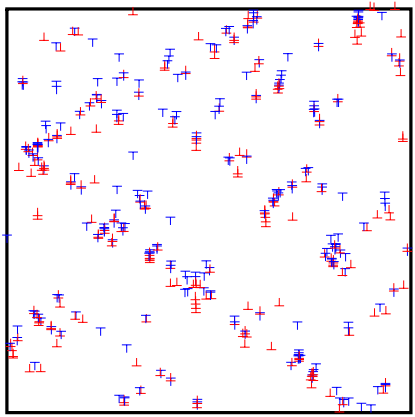}
\end{center}
\caption{A relaxed dislocation configuration with $\epsilon_\text{a}=5\cdot10^{-2}L/b$, with $L$ being the system size (anharmonic case). \label{fig:anharm}}
\end{figure}

As it can be seen in figures \ref{fig:harm} and \ref{fig:anharm}, in the harmonic case the relaxed configuration consists of randomly placed dipoles and short walls, while the presence of the extra interaction term leads to the formation of dense multipoles arranged in a structure with some periodic character \cite{groma_anharm}.

\section{Dislocation--solute atom interaction}

Solute atoms can strongly modify the collective properties of dislocations. Among other things they can lead to plastic instabilities (for a recent review see \cite{ananta}). In this section we show that the effect of solute atoms can be easily incorporated into the variational framework of section \ref{sec:var}. We restrict our considerations for straight edge dislocations with Burgers vectors parallel to the $x$ axis, but it is straightforward to generalise the method for 3D.

The Gibbs free energy of a coupled system can be always given as the sum of the Gibbs free energy of the two uncoupled systems and a coupling term. Therefore, if we add to the plastic free energy the Gibbs free energy contribution of the solute atoms than, after extremisation in the stress function, we shall arrive at the free energy of the dislocation--solute system. According to equation (\ref{eq_Echi}), the plastic free energy of the parallel edge dislocation system considered is
\begin{equation}
	P_\text{d}[\chi,\kappa]= \int \left[-\frac{1-\nu}{4\mu}(\triangle \chi)^2+b\chi(\partial_2 \kappa)\right] \text{d}^2r, \label{eq_Fd}
\end{equation}
where $\kappa$ is the signed dislocation density (geometrically necessary density, GND), defined as $\kappa:=\alpha_{31}/b$, that is the only non-vanishing component of the dislocation density tensor for the dislocation geometry considered.

For the solute atoms we assume that their concentration $c$ is close to the equilibrium concentration $c_{\infty}$. In this case the Gibbs free energy of the solute atoms can be given with the quadratic form
\begin{equation}
	G_\text{c}[c]:=\int \alpha (c-c_{\infty})^2 \text{d}^2r,
\end{equation}
where $\alpha$ is a constant (which may depend on $c_{\infty}$).

To determine the form of the coupling term we use the well-known fact that beside concentration gradient, the pressure gradient also causes solute atom diffusion. According to the principles of irreversible thermodynamics, the solute atom current is proportional to the gradient of the chemical potential $\mu_c=\delta G/\delta c$.  So, taking the coupling term in the form
\begin{equation}
	G_{\rm coupling}[\chi,c]:=\int \beta c p \, \text{d}^2r,
\end{equation}
where $\beta$ is constant, the total plastic free energy
\begin{equation}
\begin{split}
	P[\chi,\kappa,c]&=P_\text{d}[\chi,\kappa]+G_\text{c}[c]+G_{\rm coupling}[\chi,c] \\
	&=\int \left[-\frac{1-\nu}{4\mu}(\triangle \chi)^2+b\chi(\partial_2 \kappa)+ \alpha (c-c_{\infty})^2-\beta c \triangle \chi \right ] \text{d}^2r \label{c_kappa}
\end{split}
\end{equation}
obviously gives the right free energy together with the expected form of $\mu_c$. It should be mentioned that (\ref{c_kappa}) results in linear equations for the stress and the solute atom concentration. Non-linearity can also be treated within the framework proposed, but it is out of the scope of the paper.

Since this is mostly a demonstration of the way the variational approach works, we now restrict our analysis only to static problems. Dynamical aspects of the interaction between dislocations and solute atoms will be considered in forthcoming publications. One can find from the equilibrium conditions $\delta P/\delta \chi=0$ and $\delta P/\delta c=0$ that
\begin{eqnarray}
	\frac{1-\nu}{2\mu}\triangle^2\chi+\beta\triangle c =b \, \partial_2 \kappa
\end{eqnarray}
and
\begin{eqnarray}
	\alpha(c-c_{\infty})=\beta \triangle \chi.
\end{eqnarray}
By combining the two equations we get that
\begin{eqnarray}
	\left[\frac{1-\nu}{2\mu}+\frac{\beta^2}{\alpha}\right]\triangle^2\chi=b \partial_2 \kappa.
\end{eqnarray}
A remarkable feature of the above equation is that, apart from a constant multiplier, the functional form of the stress function $\chi$ is not effected by the solute atoms. Moreover, the solute atom concentration is proportional to the pressure caused by the dislocations. Certainly, the result obtained is not new, it is the well-known Cottrell atmosphere of solute atoms around dislocation lines \cite{cottrell}. But it illustrates very well the fact that the coupled system of dislocations and solute atoms can be treated with the variational framework suggested in section \ref{sec:var}. However, it is important to mention that the collective dynamics of the coupled system can also be studied with the variational method. This will be discussed in a forthcoming paper.

\section{Conclusions}

A variational framework has been presented to determine the stress field and the interaction energy of dislocation systems. It is demonstrated that the method allows to study core regularisation problems, the role of anharmonicity and dislocation--solute atom interaction. Furthermore, Groma \emph{et al.}~have discussed in detail how the dynamics of coarse grained dislocation densities can be obtained from an effective variational free energy \cite{groma_philmag}. This recognition led to the theory of Debye-like screening of dislocations in equilibrium \cite{groma6} and during relaxation \cite{ispanovity}. Certainly, like in other fields of physics, one can obtain the same results without using the variational approach, but this offers a systematic way for the treatment of a wide range of different problems. It should be also mentioned that other variational principles can also be constructed, which give the same result \cite{groger}. The advantage of the one presented here is that the variational functional is minimised with respect to the stress potential. So, it gives equations directly for the stress function from which the stress can be calculated directly.

\section*{Acknowledgement}

Financial supports of the Hungarian Scientific Research Fund (OTKA) under Contract No.\ K~67778 and by the Swiss FNRS are gratefully acknowledged.

\appendix
\section*{\appendixname. Calculation of the anharmonic potential}

As it is explained in section 5, elastic anharmonicity leads to the extra dislocation--dislocation interaction energy
(\ref{V1}):
\begin{equation}
	V_1(\bm r) = - \Lambda \int \frac{y-y'}{|\bm r-\bm r'|^2} \frac{y'^2}{|\bm r'|^4} {\mathrm d}^2 r',
\label{eqn:V1_a}
\end{equation}
where
\begin{equation}
	\Lambda := \frac{(b_1-b_2)b^2}{(2\pi)^3D}. \label{eqn:Lambda}
\end{equation}

The integral can be evaluated analytically leading to equation (\ref{V1_f}), now we give the derivation.

Let us first analyse the asymptotes of the integrand. At infinity it decays like $1/r'^3$ and near $\bm r \neq {\bm 0}$ diverges like $1/|\bm r' -\bm r|$, so in these cases integration gives finite result. Contrastively, near the origin the integrand is proportional to $1/r'^2$, so a lower cut-off $r_0$ has to be introduced and the integral will exhibit a logarithmic singularity. This is an artificial singularity, however, because it comes from the core region, where interactions are regular in reality. The proper treatment would be by core-regularised anharmonic analysis, which is relegated to later work.

To proceed with the calculation, we notice that (\ref{eqn:V1_a}) is the convolution of two functions, which can be rewritten as follows
\begin{align}
	-\frac{y^2}{r^4} &= \frac12 \left[ \frac{\partial^2}{\partial y^2} \ln\left( r\right) - \frac{1}{r^2} \right], \\
	\frac{y}{r^2} &= \frac{\partial}{\partial y} \ln\left( r\right),
\end{align}
where the lower cut-off $r_0$ is omitted from the logarithms but will be reinstated in the end. Substitution into (\ref{eqn:V1_a}) yields
\begin{equation}
	V_1(\bm r) = \frac{\Lambda}{2} \left[ V_{11}(\bm r) + V_{12}(\bm r)\right], \label{eqn:V1_sep}
\end{equation}
with
\begin{align}
	V_{11}(\bm r) & := \partial_y \ln(r) \otimes \partial_y^2 \ln(r) =\ln(r) \otimes \partial_y^3 \ln(r), \label{eqn:V11} \\
	V_{12}(\bm r) & :=- \partial_y \ln(r) \otimes r^{-2} =- \ln(r) \otimes \partial_yr^{-2}, \label{eqn:V12}
\end{align}
where $\otimes$ denotes convolution:
\begin{equation}
	f (\bm r)\otimes g(\bm r) := \int \mathrm{d}^2 r' \, f(\bm r-\bm r') g(\bm r').
\end{equation}
We then recall that the radial logarithm is the Green function of the Poisson's equation in the plane, so a convolution like $\phi(\bm r)= \ln(r) \otimes g(\bm r)$ is just the solution of $ \triangle \phi(\bm r) = 2\pi g(\bm r)$. Therefore, to calculate (\ref{eqn:V11}) and (\ref{eqn:V12}), we have to solve the Poisson's equation with the inhomogeneous terms $\partial_y^3 \ln(r)$ and $-\partial_yr^{-2}$. But since those functions are derivatives of radial functions, it suffices to solve the radial Poisson's equation and differentiate the solution later. The radial solutions needed for our considerations are
\begin{align}
 	g(\bm r) & = \ln(r) &\Rightarrow \qquad \qquad &\phi(\bm r) = \frac{\pi}{2} r^2 \left[ \ln(r)-1 \right], \\
	g(\bm r) & = -r^{-2} &\Rightarrow \qquad \qquad & \phi(\bm r) = -\pi \ln^2(r),
\end{align}
whence we get
\begin{align}
	V_{11}(\bm r) = \frac{\pi}{2} \partial_y^3 r^2 \ln(r), \label{eqn:V11b} \\
	V_{12}(\bm r) = - \pi \partial_y \ln^2(r). \label{eqn:V12b}
\end{align}
Reintroducing the cut-off $r_0$ into the logarithms and performing the differentiation gives by (\ref{eqn:Lambda}) and (\ref{eqn:V1_sep}) the leading anharmonic term to the dislocation--dislocation interaction energy:
\begin{equation}
	V_1(\bm r) = \frac{(b_1-b_2)b^2}{16\pi^2 D} \left[ \frac{y(3x^2+y^2)}{r^4} -2 \frac{y}{r^2} \ln\left( \frac{r}{r_0} \right) \right].
\label{eqn:V1_final_a}
\end{equation}

\end{document}